\documentclass[twocolumn,showpacs,preprintnumbers,amsmath,amssymb]{revtex4}

\usepackage{graphicx}
\usepackage{dcolumn}
\usepackage{bm}
\usepackage{epsfig}
\usepackage{longtable} 

\begin{document}

\preprint{}

\title{
Half-life of the electron-capture decay of $^{97}$Ru:  Precision measurement shows
no temperature dependence
}

\author{J.R. Goodwin}

\author{V.V. Golovko}

\altaffiliation {Present address: Department of Physics, Queen's
University, Stirling Hall, Kingston, ON, Canada K7L3N6}

\author{V.E. Iacob}

\author{J.C. Hardy}
\email{hardy@comp.tamu.edu}

\affiliation{ Cyclotron Institute, Texas A\&M University, College Station, Texas 77843, USA}
\homepage{http://cyclotron.tamu.edu/}

\date{\today}

\begin{abstract}
We have measured the half-life of the electron-capture (ec) decay of
$^{97}$Ru in a metallic environment, both at low temperature (19K), and
also at room temperature.  We find the half-lives at both temperatures to
be the same within 0.1\%.  This demonstrates that a recent claim that the
ec decay half-life for $^7$Be changes by $0.9\% \pm 0.2\%$ under similar circumstances certainly cannot be generalized to other ec decays.  Our
results for the half-life of $^{97}$Ru, 2.8370(14)\,d at room temperature
and 2.8382(14)\,d at 19K, are consistent with, but much more precise than,
previous room-temperature measurements.  In addition, we have also
measured the half-lives of the $\beta^-$-emitters $^{103}$Ru and
$^{105}$Rh at both temperatures, and found them also to be unchanged.
 
\end{abstract}

\pacs{}

\maketitle

\section{\label{sec:introd} INTRODUCTION}

Since the early days of nuclear science, nearly a century ago, it has been widely accepted that the decay constants of radioactive isotopes decaying by $\alpha$, $\beta^-$ or $\beta^+$ emission are independent of all physical or chemical conditions such as pressure, temperature and material surroundings.  This belief was based on numerous measurements in the early 1900's, some of which claimed remarkable precision (see \cite{Em72} for an interesting review): for example, Curie and Kamerlingh Onnes \cite{Cu13} in 1913 determined that the decay constant of a radium preparation did not change by more than 0.1\% when cooled to 20K.  In contrast, decays proceeding by internal conversion or electron capture (ec), to which atomic electrons contribute directly, were placed in a different category, being potentially susceptible to their chemical --- though not physical --- condition.  There is a long history of $^7$Be decay measurements that demonstrate small but detectable effects on that isotope's decay constant caused by its chemical environment.

Quite recently, however, measurements have been reported claiming relatively large changes in half lives for $\alpha$, $\beta^-$, $\beta^+$ and ec decays depending on whether the radioactive parent was placed in an insulating or conducting host material, and whether the latter was at room temperature or cooled to 12K.  Specifically, $^{210}$Po, an $\alpha$ emitter, when implanted in copper was reported to exhibit a half life shorter by 6.3(14)\% at 12K than at room temperature \cite{Ra07}; the $\beta^-$ emitter $^{198}$Au in a gold host reportedly had a half-life longer by 3.6(10)\% at 12K \cite{Sp07}; $^{22}$Na, which decays predominantly (90\%) by $\beta^+$ emission, was measured as having a 1.2(2)\% shorter half life at 12K \cite{Li06}; and $^7$Be, which decays by pure electron capture, apparently had a half-life longer by 0.9(2)\% at 12K in palladium and by 0.7(2)\% in indium \cite{Wa06}.  The authors of these reports also proposed a theoretical explanation of their observations based on quasi-free electrons --- a ``Debye plasma" --- causing an enhanced screening effect in metallic hosts.  This would lead to host-dependent half-lives and a smooth dependence of half-life on temperature in a metal.

Needless to say, these claims led to considerable popular interest, not least because they could potentially have contributed to the improved disposal of radioactive waste \cite{Mu06}.  Not remarked on at the time, though, was the impact that such a result would also have on all half-lives that have ever been quoted with sub-percent precision.  Of greatest concern to us were the half-lives of superallowed $0^+$$\rightarrow$$0^+$ $\beta^+$ transitions, essential to fundamental tests of the Standard Model \cite{Ha09}.  Their precision has typically been quoted to less than 0.05\%, well below the temperature and host-material dependence claimed by the new measurements \cite{Ra07,Sp07,Li06,Wa06}.

Based on this concern, we first repeated the measurement on the decay of $^{198}$Au ($t_{1/2}$ = 2.7\,d) in gold \cite{Go07}. While the original measurement by Spillane et al. \cite{Sp07} followed the decay for only a little over one half-life, we recorded the decay with much better statistics for over 10 half-lives at both room temperature and at 19K.  Our results showed the half-lives at the two temperatures to be the same within 0.04\%, a limit two orders of magnitude less than the difference claimed by Spillane et al.  This null result was subsequently confirmed by two other measurements of $^{198}$Au, which set limits of 0.13\% in a Al-Au alloy host \cite{Ku08} and 0.03\% in gold \cite{Ru08}.  The latter reference also reported a new $^{22}$Na decay measurement, which set an upper limit on the temperature dependence of that $\beta^+$ decay at 0.04\%, again nearly two orders of magnitude below the earlier claim, in this case by Limata et al. \cite{Li06}.  For $\alpha$ decay, the $^{210}$Po measurement has not yet been repeated but low-temperature measurements on a variety of other $\alpha$ emitters \cite{St07,Se07} have set upper limits of 1\% on any possible temperature dependence in those cases.  Though significantly lower than the temperature dependence claimed to have been observed in reference \cite{Ra07}, this 1\% limit is considerably less stringent than the limits obtained for $\beta^-$ and $\beta^+$ decays.

The status of electron-capture decay is also less definitive.  One new measurement of $^7$Be decay in copper \cite{Ku08} found no temperature dependence greater than 0.3\% but another \cite{Ni07} actually found a small change in half-life --- 0.22(8)\% --- depending on whether the host material was a conductor (Cu or Al) or an insulator (Al$_2$O$_3$ or PVC), both at room temperature.  In neither case is the result as precise as has been achieved for $\beta^-$ and $\beta^+$ decays.  Furthermore, since $^7$Be is known to show effects from its chemical environment, it is difficult to be certain about the cause of any observed effect and even more difficult to generalize its behavior to the electron-capture decay of other nuclei for which the $K$-shell electrons are much better shielded from the external environment.

\begin{figure}[t]
\epsfig{file=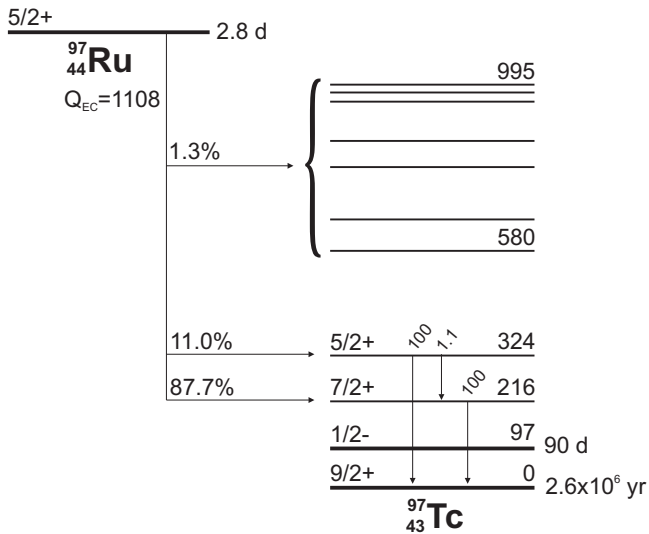,width=8.5cm}
\caption{Partial scheme for the electron-capture decay of $^{97}$Ru, showing the dominant two transitions and the $\gamma$ rays that follow them.  The information is taken from Ref. \cite{Au03}.  We measured the $^{97}$Ru half-life by following the time decay of the 216-keV $\gamma$ ray.}
\label{fig1}
\end{figure}

We thus set out to determine the temperature dependence for the ec-decay half-life of a nucleus with a $Z$ that is considerably larger than that of $^7$Be.  Our goal was to achieve a precision comparable to that obtained for $\beta^-$ and $\beta^+$ decays, {\it i.e.}\,$\leq$0.1\%.  For our measurement we sought a nucleus that decays entirely by electron capture with a few-day half-life and a delayed $\gamma$ ray that can be cleanly detected.  It also had to be producible by thermal-neutron activation so that we could obtain statistically useful quantities without serious contaminants.  Although there are not a lot of candidates to choose among, we found $^{97}$Ru satisfied all our conditions.  Its decay scheme appears in Fig.~\ref{fig1}.  We report here measurements of the half-life of $^{97}$Ru at room temperature and at 19K as measured via its 216-keV $\beta$-delayed $\gamma$ ray.  We have found no temperature dependence in the results. Our upper limit is 0.1\%, an order of magnitude below the effect claimed for $^7$Be \cite{Wa06}.

\section{\label{sec:Appar} Apparatus and set-up}

We used the same set-up for both the cold and room-temperature measurements. As we did previously for our $^{198}$Au half-life measurement \cite{Go07}, we placed the ruthenium sample between two copper washers and fastened the assembly directly onto the cold head of a CryoTorr7 cryopump with four symmetrically placed screws. A 70\% HPGe detector was placed facing the sample on the cryopump axis, just outside the pump's coverplate, into which a cavity had been bored so that only 3.5 mm of stainless steel stood between the detector face and the sample.  The total distance between the detector face and the ruthenium sample was 49 mm and remained unchanged throughout the experiment. We monitored the temperature of the sample with a temperature-calibrated silicon diode (Lakeshore Cryogenics DT-670) \cite{LS} fastened in the same way as the ruthenium sample and placed right next to it on the head itself.  The diode was connected to a Lakeshore Model 211 temperature monitor. 

For the low-temperature measurement, we first used a roughing pump to bring the pressure down to about 9 mtorr, and then switched on the cryopump.  Although the cold head, where the sample was located, is nominally expected to reach 12K, we measured its temperature to be between 18.2 and 20.8K, with an average value of 19K.  The arrangement for the room temperature measurement was identical except that these pumping and cooling steps were omitted.  Note that we did not alternate temperatures for a single source but rather made a complete decay measurement at one temperature with one source at a fixed geometry; then, with a fresh source, we made a similar dedicated measurement at the other temperature.  Thus our results are entirely independent of any geometrical or source differences that might have occurred between the two measurements.

Our sample was a single crystal in the form of a circular disc, 8 mm in diameter and 1 mm thick, obtained from Goodfellow Corp. According to the supplier, the chemical purity of the material was 99.999\%, with no identifiable impurities. For each measurement, the metal crystal was initially activated for 10 seconds in a flux of $10^{13}$ neutrons/cm$^2$\,s, at the Texas A\&M Triga reactor. This activated crystal was then fastened directly to the cold head of the cryopump, ensuring a good thermal contact over the whole crystal area.

\begin{figure*}[t]
\epsfig{file=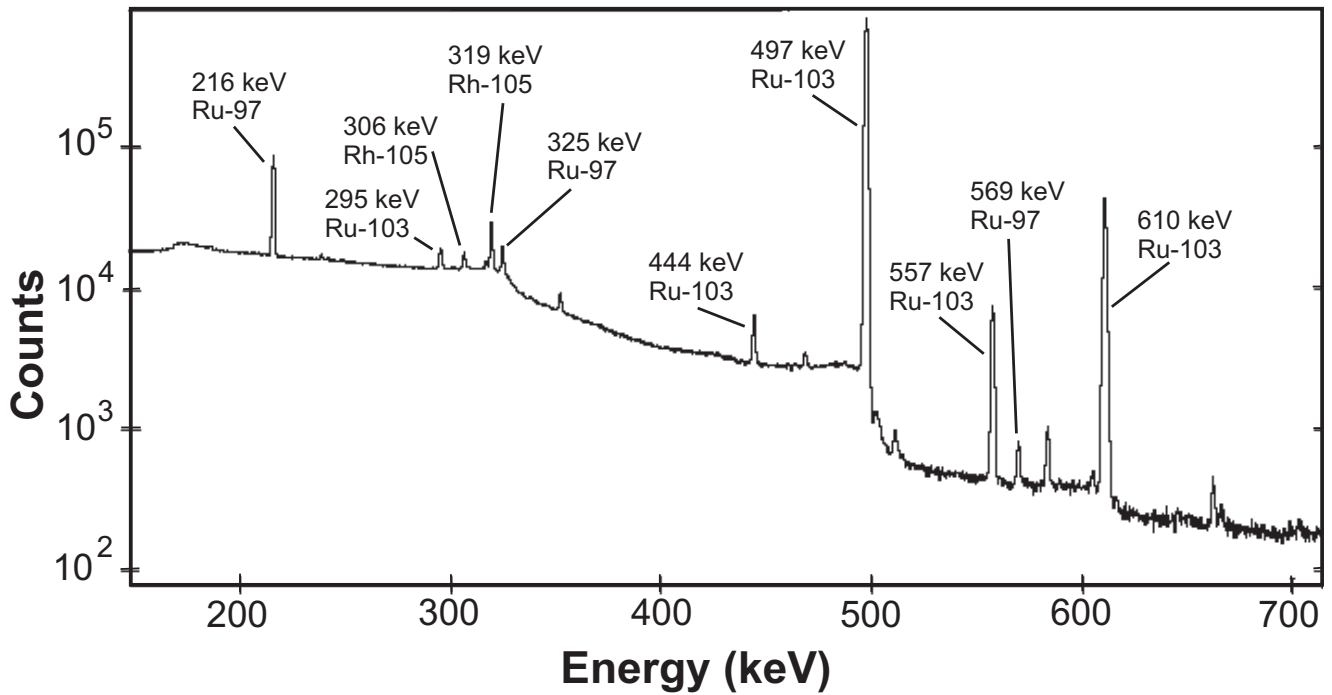,width=17.5cm}
\caption{The principal peaks from the decays of $^{97}$Ru, $^{105}$Rh, and $^{103}$Ru are marked both by the parent isotope and the energy in keV.  These are all pure peaks with the following two exceptions: the peak at 295 keV consists of approximately 75\% $^{103}$Ru and 25\% $^{226}$Ra; and the peak at 610 keV consists of approximately 95\% $^{103}$Ru and 5\% $^{214}$Bi.  The remaining unmarked peaks are well-known background peaks identified in a separate background measurement.}
\label{fig2}
\end{figure*}

For the measurement itself, sequential $\gamma$-ray spectra were recorded from the HPGe detector. The detector signals were amplified and sent to an analog-to-digital converter, which was an Ortec TRUMP$^{\rm TM}$-8k/2k card \cite{OR} controlled by Maestro software, which was installed on a PC operating under Windows-XP.  During the entire period of the measurements, our computer clock was synchronized daily against the signal broadcast by WWVB, the radio station operated by the U.S. National Institute of Standards and Technology.  For both the room- and low-temperature measurements, six-hour spectra were acquired sequentially for approximately one month.  In each case, more than 110 $\gamma$-ray spectra were recorded. 

The TRUMP$^{\rm TM}$ card uses the Gedcke-Hale method \cite{Je81} to correct for dead-time losses.  By keeping our system dead time below about 4\% and recording all our spectra for an identical pre-set live time, we ensured that our results were essentially independent of dead-time losses. However, at a precision level of 0.1\% or better, pile-up can also become an issue, so we carefully tested our system for residual rate-dependent effects, as reported in our previous article on $^{198}$Au \cite{Go07}.  We first measured the 662-keV $\gamma$-ray peak from a $^{137}$Cs source alone, and then remeasured that source a number of times in the presence of a $^{133}$Ba source, which was moved closer and closer to the detector in order to increase the dead time and the number of chance coincidences. Each measurement was made for the same pre-set live time. We then obtained from each measurement the number of counts in the 662-keV peak and, from the decrease in that number as a function of increasing dead time, we determined that the fractional residual loss amounted to $5.5(2.5) \times 10^{-4}$ per 1\% increase in dead time. At the count rates experienced during our $^{97}$Ru measurements, the required correction was never greater than 0.2\% but it was nevertheless applied to all spectra.

\section{\label{sec:Res} Results and Analysis}

A typical $\gamma$-ray spectrum, one of the more than 220 obtained, is shown in Fig.~\ref{fig2}.  Apart from the weak peaks due to room background, the only observed $\gamma$ rays are from the decays of $^{97}$Ru ($t_{1/2}$ = 2.8\,d), $^{103}$Ru (39\,d) and $^{105}$Rh (35\,h); the latter is the daughter of $^{105}$Ru (4.4\,h), which had already decayed away by the time this spectrum was recorded.  The appearance of these three ruthenium isotopes is consistent with their being produced by neutron activation of naturally occurring ruthenium.  The 216-keV $\gamma$-ray peak from $^{97}$Ru is seen to be clear of any other peaks and to lie on a smooth, though rather high, background.

\begin{figure}[t]
\epsfig{file=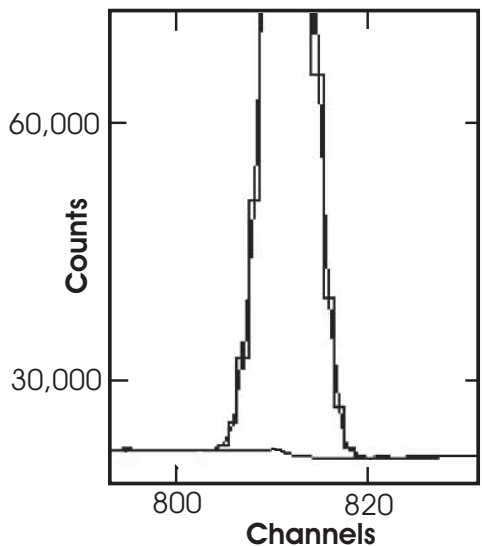,width=6.3cm}
\caption{Example of a measured 216-keV $\gamma$-ray peak together with the fit obtained from GF3. Note that the vertical scale has been greatly expanded to display the low-level background, and the quality of the fit to it.  This spectrum was taken about five days after counting began; the peak contained a net of about 600,000 counts. }
\label{fig3}
\end{figure}

The 216-keV $\gamma$-ray peak in each recorded spectrum was analyzed with GF3, a least-square peak-fitting program in the RADware series \cite{Rapc}.  This program allowed us to be very specific in determining the correct background for a peak, and the 216-keV peak in each spectrum was visually inspected to this end.  So far as possible, the same criteria were applied to each spectrum. Fig.~\ref{fig3} shows a sample peak and the fitted background, from which its area was determined.  

\begin{figure}[b]
\epsfig{file=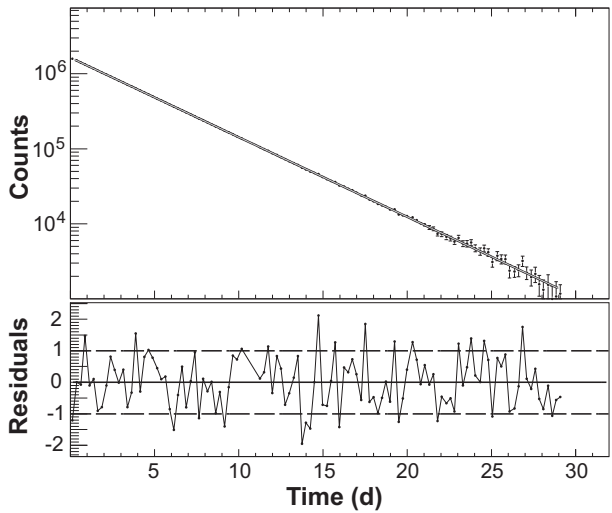,width=8cm}
\caption{Decay of $^{97}$Ru in ruthenium metal, at room temperature. Experimental data appear as dots; the straight line is a fit to these data. Normalized residuals appear at the bottom of the figure.  The dashed lines in the residuals plot represent $\pm$1 standard deviation from the fitted value.}
\label{fig4}
\end{figure}

\begin{figure}[b]
\hspace{1mm}
\epsfig{file=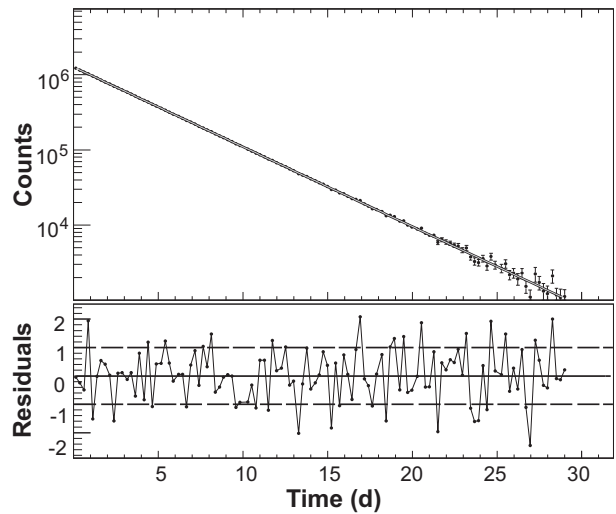,width=8.3cm}
\caption{Decay of $^{97}$Ru in ruthenium metal, at 19K. Experimental data appear as dots; the straight line is a fit to these data.  Normalized residuals appear at the bottom of the figure.  The dashed lines in the residuals plot represent $\pm$1 standard deviation from the fitted value.}
\label{fig5}
\end{figure}

In total, 229 spectra were subjected to this careful analysis, and the counts recorded in the 216-keV peak for each were corrected for residual losses (see Sect.~\ref{sec:Appar}). The results for the room temperature and 19K measurements are plotted as a function of time in Figs.~\ref{fig4} and \ref{fig5}. The decay curves were then analyzed by a maximum-likelihood fit to a single exponential. The code we used, which is based on ROOT \cite{Br97}, has previously been tested by us to 0.01\% precision with Monte Carlo generated data.  The data in Figs.~\ref{fig4} and \ref{fig5} yield $^{97}$Ru half-lives (with statistical uncertainties only) of 2.8370(13)\,d for the room temperature measurement, and 2.8382(13)\,d for the one at 19K. The difference between these two results is 0.0012(18)\,d, which gives an upper limit of 0.0030\,d, or 0.1\%, on any temperature-dependent difference at the 68\% confidence level.

\begin{figure}[t]
\epsfig{file=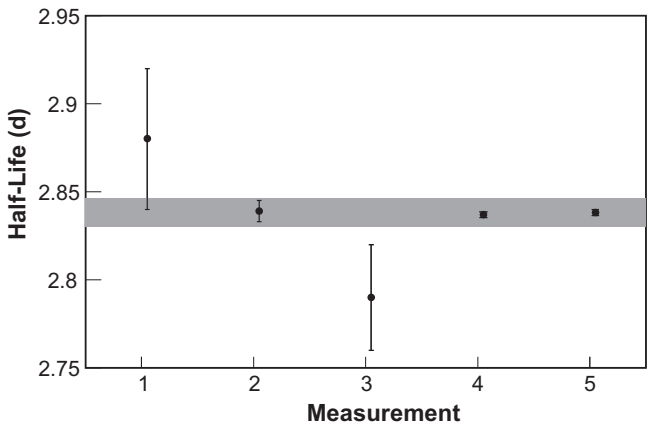,width=8.5cm}
\caption{The data points represent all published measurements of the $^{97}$Ru half-life that have been quoted with better than 2\% precision.  The results are plotted in chronological order from left to right, measurements 1--3 being those of Katcoff {\it et al.}\,\cite{Ka58}, Silvester {\it et al.}\,\cite{Si79} and Kobayashi {\it et al.}\,\cite{Ko98}, respectively; the shaded area represents the weighted average of these measurements.  Measurements 4 and 5 are the room-temperature and 19K results of the present measurement.  The data in the figure are taken from those tabulated in Table~\ref{table1}.}
\label{fig6}
\end{figure}

The half-life values taken from the computer fits incorporate the correction for residual losses described in Sect.~\ref{sec:Appar}, but they do not yet include the uncertainty in that correction, since it is correlated for the two measurements and does not contribute to the difference between them.  However, for our measurements to be compared with previous measurements of the $^{97}$Ru half-life, this systematic uncertainty is now incorporated, and yields the results 2.8370(14)\,d and 2.8382(14)\,d for the room temperature and 19K measurements, respectively.  These values are compared with previous measurements of the $^{97}$Ru half-life in Table~\ref{table1} and Fig.~\ref{fig6}, where it can be seen that our results at both temperatures are much more precise than, but are entirely consistent with, the previous ones, all of which were presumably made at room temperature. 

As a byproduct of our primary measurement on $^{97}$Ru we have also extracted from the same spectra half-lives at both temperatures for the nuclides $^{103}$Ru and $^{105}$Rh, both $\beta^-$ emitters.  For $^{103}$Ru, we monitored the 497-keV peak in all 237 spectra, while for the shorter lived $^{105}$Rh there were only sufficient statistics for us to use 100 spectra to follow the 319-keV peak (see Fig.~\ref{fig2}).  These peaks were subjected to the same meticulous examination, fitting and analysis as just described for the 216-keV peak of ${97}$Ru.  Incorporating only statistical uncertainties, we obtained half-life values for $^{103}$Ru, of 39.210(16)\,d at room temperature, and 39.219(25)\,d at 19K, which are statistically the same within 0.1\%.  For $^{105}$Rh, our half-life values with statistical uncertainties only are 35.357(36)\,h at room temperature and 35.319(23)\,h at 19K, again the same, but in this case within 0.2\%. 

\begin{table}[b]
\begin{center}
\caption{\label{table1} Measurements of the $^{97}$Ru half-life made since 1946.}
\vskip 1mm
\begin{ruledtabular}
\begin{tabular}{lll}
 & & \\[-3mm]
Half-life (d) & Reference & Year \\
& & \\[-3mm]
\hline
 & & \\[-2mm]
2.8(3) & Sullivan {\it et al.} \cite{Su46}  &  1946  \\
2.8(1) & Mock {\it et al.} \cite{Mo48} &  1948  \\
2.88(4) & Katcoff {\it et al.} \cite{Ka58} & 1958  \\
2.9(1) & Cretzu {\it et al.} \cite{Cr66} & 1966  \\
2.839(6) & Silvester {\it et al.} \cite{Si79} & 1979  \\
2.79(3) & Kobayashi {\it et al.} \cite{Ko98} & 1998  \\
 & & \\[-2mm]
2.838(6) & Weighted average \\
 & & \\[-2mm]
\multicolumn{2}{l}{This measurement:} & \\
2.8370(14) & Room temperature & 2009 \\
2.8382(14) & 19K & 2009  \\
0.0012(18) & Difference & \\
\end{tabular}
\end{ruledtabular}
\end{center}
\end{table}

As we did when making the temperature comparison with $^{97}$Ru, we have so far quoted half-life values for $^{103}$Ru and $^{105}$Rh that do not yet include the (correlated) uncertainty attributable to residual losses (see Sect.~\ref{sec:Appar}).  We include that now in order to compare our results with previous half-life measurements.  Our final half-life results for $^{103}$Ru then become 39.210(38)\,d at room temperature, and 39.219(35)\,d at 19K; and for $^{105}$Rh our results are 35.357(37)\,h at room temperature and 35.319(24)\,h at 19K.  Note that the effect of residual losses on the uncertainty of the $^{103}$Ru half-life is much greater than it is for the $^{105}$Rh half-life.  Since $^{103}$Ru is much longer lived, our data only encompass a little more than one half-life, during which time the overall count-rate in our detector has decreased significantly.

Unlike the situation for the other two radionuclides studied in this work, the half-life of $^{103}$Ru has been measured rather precisely in the past, with four of the previous results being of comparable precision to our current ones.  Unfortunately, though, the earlier results are not particularly consistent with one another, as can be seen in Table~\ref{table2}.  The normalized $\chi^2$ for the average of all previous measurements is 3.0, which results in our scaling up the uncertainty assigned to that average by a factor of 1.7.  In comparison with this average value, our results are slightly low, though the discrepancy is not statistically very significant.  Note also that our results are completely consistent with the 1981 value obtained by Miyahara et al. \cite{Mi81}.

\begin{table}[b]
\begin{center}
\caption{\label{table2} Measurements of the $^{103}$Ru half-life quoted with sub-percent precision.}
\vskip 1mm
\begin{ruledtabular}
\begin{tabular}{lll}
 & & \\[-3mm]
Half-life (d) & Reference & Year \\
& & \\[-3mm]
\hline
 & & \\[-2mm]
39.5(3) & Flynn {\it et al.} \cite{Fl65}  &  1965  \\
39.35(5) & Debertin \cite{De71} &  1971  \\
39.254(8) & Houtermanns {\it et al.} \cite{Ho80} & 1980  \\
39.214(13) & Miyahara {\it et al.} \cite{Mi81} & 1981  \\
39.260(20) & Vaninbroukx {\it et al.} \cite{Va81} & 1981  \\
39.272(16) & Walz {\it et al.} \cite{Wa83} & 1983  \\
 & & \\[-2mm]
39.250(10) & Weighted average (scale factor, 1.7) \\
 & & \\[-2mm]
\multicolumn{2}{l}{This measurement:} & \\
39.210(38) & Room temperature & 2009 \\
39.219(35) & 19K & 2009  \\
0.009(30) & Difference & \\
\end{tabular}
\end{ruledtabular}
\end{center}
\end{table}

There are only three previous measurements of the $^{105}$Rh half-life, none more recent than 1967; they are listed in Table~\ref{table3}. Strikingly, the earliest measurement \cite{Br62} has the tightest, $\pm$0.06\%, uncertainty and a half-life value that disagrees completely with the two later measurements.  The weighted average of all three measurements yields a normalized $\chi^2$ of 22 and, as shown in Table~\ref{table3}, its uncertainty consequently requires scaling by a factor of 4.7.  Under the circumstances, it seems more reasonable not to use this average value, but simply to disregard the offending measurement and average the two remaining, mutually consistent, results \cite{Pi65,Ko67}.  When compared with this new average, our results are a factor of two more precise and lie slightly lower.  Considering that even the two previous measurements that have been retained are more than 40 years old and that the difference between their average and our recent results is less than two standard deviations, there seems little reason for concern.

\begin{table}
\begin{center}
\caption{\label{table3} Measurements of the $^{105}$Rh half-life.}
\vskip 1mm
\begin{ruledtabular}
\begin{tabular}{lll}
 & & \\[-3mm]
Half-life (h) & Reference & Year \\
& & \\[-3mm]
\hline
 & & \\[-2mm]
35.88(2) & Brandhorst and Cobble \cite{Br62}  &  1962  \\
35.4(1) & Pierson \cite{Pi65} &  1965  \\
35.47(8) & Kobayashi \cite{Ko67} & 1967  \\
 & & \\[-2mm]
35.84(9) & Weighted average (scale factor, 4.7) \\
35.44(6) & Weighted average of \cite{Pi65} and \cite{Ko67} \\
 & & \\[-2mm]
\multicolumn{2}{l}{This measurement:} & \\
35.357(37) & Room temperature & 2009 \\
35.319(24) & 19K & 2009  \\
0.038(43) & Difference & \\
\end{tabular}
\end{ruledtabular}
\end{center}
\end{table}

\section{\label{sec:Conc} Conclusions }

We have measured the half-life of $^{97}$Ru in ruthenium metal at room temperature and at 19K, and have found the results to be the same within 0.1\%. Since the maximum decay energy for any allowed transition from $^{97}$Ru is 892 keV, the nucleus must decay by pure electron capture.  Three years ago, Wang {\it et al.}\,\cite{Wa06} reported half-life measurements of another pure electron-capture emitter, $^7$Be, situated in both palladium and indium metals, in which they observed differences of 0.9(2)\% and 0.7(2)\%, respectively, between room temperature and 12K.  The same group also reported cases of temperature dependence for $\alpha$, $\beta^-$ and $\beta^+$ decay modes \cite{Ra07,Sp07,Li06} and interpreted them all as the result of a ``Debye plasma," which purportedly acts in any metal host and leads to a smooth dependence of half-lives on temperature.  In that context, their result for $^7$Be decay was understood to be the indication of a generic property of all ec-decays rather than a unique property of $^7$Be.

Obviously we cannot comment on the validity of the $^7$Be measurement itself, but we can certainly refute any suggestion that the half-lives of ec-decays in general exhibit significant temperature dependence when the source is placed in a metal host.  Wang {\it et al.}\,\cite{Wa06} used their model to calculate that the half-life of $^7$Be in a metal should change by 1.1\% between T = 293 and 12K, a result that agrees reasonably well with their measured values.  Using the same model, we calculate that the half-life change for the $^{97}$Ru decay should be 11.2\% between T = 293 and 12K and 8.4\% between T = 293 and 19K, the temperature we obtained.  Our measured upper limit on any half-life change over this temperature range is nearly two orders of magnitude less than this model prediction.  We have previously demonstrated that the "Debye model" has no validity for $\beta^-$ decay \cite{Go07}; we can now state with equal confidence that it also does not apply to ec decay.

As a byproduct of this primary measurement, we also obtained half-life data for two $\beta^-$ emitters, $^{103}$Ru and $^{105}$Rh, at room temperature and 19K.  These results, though slightly less precise than our measurements on the $\beta^-$ decay of $^{198}$Au \cite{Go07}, nevertheless confirm our previous conclusion for that decay mode.  With any temperature dependence for $\beta^+$ decay also now ruled out at the 0.04\% level \cite{Ru08}, it has become clear that there is no reason to doubt the accuracy of nuclear weak-decay half-lives that have been quoted over the past decades with sub-percent precision and without accounting for the host material or temperature.  As has always been believed, those parameters indeed do not affect the result, at least not above the 0.1\% level. 

In all three cases, $^{97}$Ru, $^{103}$Ru and $^{105}$Rh, our measured half-lives are consistent with, and in two cases are substantially more precise than, previous measurements.

\begin{acknowledgments}

     We thank Prof. R. Watson for the generous loan of a cryopump. This work was supported by the U.S. Department of Energy under Grant No. DE-FG03-93ER40773 and by the Robert A. Welch Foundation under Grant No. A-1397.  

\end{acknowledgments}

\end{document}